\documentclass[prd,aps,twocolumn,nofootinbib,showpacs,amsmath,superscriptaddress,preprintnumbers]{revtex4-1}
\usepackage{graphicx}
\usepackage{epstopdf}
\usepackage{bm}
\usepackage{textcomp}
\usepackage{epsfig}
\usepackage{graphics}
\usepackage{xspace}
\usepackage{amssymb}
\usepackage{amsmath}
\usepackage{xcolor}
\usepackage[utf8]{inputenc}
\usepackage[normalem]{ulem}

\def\OMIT#1{}

\newcommand{\nn}{\nonumber}

\newcommand{\bea}{\begin{eqnarray}}
\newcommand{\eea}{\end{eqnarray}}

\newcommand{\gsim}{\mathrel{\rlap{\lower4pt\hbox{\hskip1pt$\sim$}}\raise1pt\hbox{$>$}}}

\newcommand{\be}{\begin{equation}}
\newcommand{\ee}{\end{equation}}

\DeclareMathOperator{\Tr}{Tr}

\begin{document}
\title{Anti-$k_T$ jet function at next-to-next-to-leading order}
\author{Hao-Yu Liu}
\email{hyliu1@bnu.edu.cn}
\affiliation{Center of Advanced Quantum Studies, Department of Physics, Beijing Normal University, Beijing 100875, China}
\author{Xiaohui Liu}
\email{xiliu@bnu.edu.cn}
\affiliation{Center of Advanced Quantum Studies, Department of Physics, Beijing Normal University, Beijing 100875, China}
\affiliation{Center for High Energy Physics, Peking University, Beijing 100871, China}
\author{Sven-Olaf Moch}
\email{sven-olaf.moch@desy.de}
\affiliation{II. Institut f\"ur Theoretische Physik Universit\"at Hamburg,
Luruper Chaussee 149, D-22761 Hamburg, Germany}

\begin{abstract}
Jets constructed via clustering algorithms (e.g., anti-$k_T$, soft-drop) have 
been proposed for many precision measurements, such as the strong coupling
$\alpha_s$ and the nucleon intrinsic dynamics. However, the theoretical
accuracy is affected by missing QCD corrections at higher orders 
for the jet functions in the associated factorization theorems. Their
calculation is complicated by the jet clustering procedure. In this work, we
propose a method to evaluate jet functions at higher orders in QCD. The
calculation involves the phase space sector decomposition with suitable soft
subtractions. As a concrete example, we present the quark-jet function using
the anti-$k_T$ algorithm with E-scheme recombination at next-to-next-to-leading order.
\end{abstract}

\preprint{DESY 21-032}
\maketitle

\section{Introduction}
Jets have been the center of attention at many frontiers of high-energy
particle and nuclear physics. At the Large Hadron Collider (LHC), for many
years, jets have served as indispensable tools for precision test of the
Standard Model (SM) and new physics searches at the TeV scale. 
Measurements at the LHC and the Relativistic Heavy Ion Collider (RHIC) 
demonstrate that jets can also be unique probes of non-perturbative dynamics, 
such as collinear parton distribution functions (PDFs)~\cite{Accardi:2016ndt} as well as 
transverse momentum dependent ones (TMDPDFs)~\cite{Angeles-Martinez:2015sea}, 
the intrinsic spin of the nucleon~\cite{Boer:2014lka,Aschenauer:2016our,Adamczyk:2017wld} 
or the hot medium effects of the quark-gluon plasma~\cite{Chien:2016led,Connors:2017ptx,Qiu:2019sfj}.  
These efforts are supported by theory developments of novel jet substructures~\cite{Larkoski:2014wba,Larkoski:2017jix}, 
and enable accurate extractions of the SM parameters~\cite{Larkoski:2017jix,Kardos:2020ppl} 
and the three-dimensional tomographic images of 
nucleons~\cite{Gutierrez-Reyes:2018qez,Liu:2018trl,Liu:2020dct,Arratia:2020nxw,Kang:2020fka,Kang:2020xez,Chien:2020hzh}
out of the jets.  
All these studies will receive a further boost at the future Electron-Ion Collider (EIC)~\cite{Boer:2011fh,Accardi:2012qut}, 
including the possibility to consider polarization degrees of freedom.

The theoretical foundation of such precision measurements roots in the
factorization of the cross section $d\sigma$ with $N$ exclusive jets with
small jet radius $R$, whose schematic form is~\cite{Becher:2015hka,Becher:2016mmh,Liu:2017pbb} 
\bea\label{eq:factorization}
d\sigma = {\cal F}_a {\cal F}_b \Tr[H S_G]  \, \prod_c^N
\sum_m\Tr[J^c_m \otimes_\Omega S^c_{cs,m}] \,,
\eea
where ${\cal F}$ encodes the parton distributions of the colliding nucleons,
such as the collinear PDFs or TMDPDFs, in case transverse momentum dependence is kept.
The traces are over the color matrices. 
$H$ is the hard function that describes the hard interaction which initiates 
the process while $S_G$ includes the wide angle radiations with long
wavelength. {The jet function containing $m$ partons is given by}~\cite{Becher:2015hka,Becher:2016mmh}
\bea\label{eq:jet-m}
J_m^c &=& \int_{\sum p^{+,\perp}_i = p^{+,\perp}_J} \prod_{i=1}^m \frac{d E_i E_i^{d-3}}{2(2\pi)^{d-1}}
\nn \\
&\times& \Theta_{\text{jet  alg.}} \sum_{\rm spins} | {\cal M}_m \rangle \langle {\cal M}_m |  \,, 
\eea
{where $| {\cal M}_m \rangle$ is the amplitude for the
  collinear field with momentum $p^\mu_J \approx \frac{p^+_J}{2} n^\mu_J$,
  with $n_J$ a light-like vector along the jet axis, to split into $m$
  particles with momenta $\{p_1\,, \dots \,, p_m\}$ and 
  $\Theta_{\text{jet alg.}}$ enforces that all the $m$ collinear particles are clustered within one jet.}

The measurements of jets and jet substructures are formulated
in the angular convolution $ \otimes_\Omega$ of the energetic collinear jet
function $J_m$ and the collinear-soft function $S_{cs,m}$, where $m$ is the
energetic collinear parton multiplicity. The angular convolution
guarantees the proper inclusion of the non-global logarithms
(NGLs)~\cite{Dasgupta:2001sh} if present. 
We remark that the multi-emission feature of the jet cross section in
Eq.~(\ref{eq:factorization}) is also shared 
by the forward scattering in the small-$x$ framework~\cite{Liu:2020mpy}.  
The necessity for increasing theoretical precision for jet cross sections, 
both at fixed order perturbation theory and with resummation, 
requires the evaluation of each component in the factorization 
Eq.~(\ref{eq:factorization}) at least to next-to-next-to-leading order (NNLO).  

On the other hand, many theoretical predictions of the jet measurements are
currently based on the simplified
formula~\cite{Liu:2012sz,Liu:2013hba,Sun:2014gfa,Liu:2018trl,Liu:2018ktv,Liu:2020dct,Liu:2020jjv,Arratia:2020nxw} 
\bea\label{eq:factorizationnll}
d\sigma = {\cal F}_a {\cal F}_b \Tr[H S_G]  \,
 \prod_c^N\, J^c \, S^c_{cs}\, e^{L_{\rm ngl}} \,, 
\eea
where one decouples the angular correlation between $J_{m}$ and $S_{cs,m}$ and 
averages each function over its solid angle in Eq.~(\ref{eq:jet-m}) to have 
$J = \sum_m \langle J_m \rangle_\Omega$ and 
$S_{cs} =  \langle S_{cs,1} \rangle_\Omega$. 
As indicated by the factor $e^{L_{\rm ngl}}$, 
the leading NGLs are resummed through the exponential. 
In order to reliably justify the logarithmic accuracy of the simplified form,
a direct NNLO calculation of the factorization is also desired. 

For many processes, the hard and the soft functions are already known to two
loops or even beyond, see, e.g.~\cite{heinrich2020collider}.
However the two-loop information for the jet function is still missing 
and therefore limits the theoretical accuracy of the resummation 
to the next-to-leading logarithm (NLL).
Compared with the other components in the factorization, the 
computation of the jet function beyond next-to-leading order (NLO) is
dramatically complicated by the presence of the recursive clustering
procedure, which is the essence of the widely used algorithms for 
jets and jet substructure, such as anti-$k_T$~\cite{Cacciari_2008} with either
the E-scheme or the winner-take-all (WTA) recombination scheme~\cite{Bertolini:2013iqa, Neill:2016vbi} 
and soft-drop grooming~\cite{Larkoski:2014wba}.

In this work, we propose a method to evaluate $J_m$ or $J$ at NNLO efficiently
using sector decomposition for the phase space together with suitable designed soft subtractions.  
We demonstrate the feasibility of the method by presenting the first
anti-$k_T$ quark-jet function at ${\cal O}(\alpha_s^2)$ using E-scheme recombination as a concrete example.

\section{NLO}
We take the well-known NLO calculation~\cite{Liu:2012sz,Arratia:2020nxw} of the quark-jet function as a warm up to illustrate our approach. The NLO calculation involves evaluating the integration of the matrix element for $q_a \to q_i g_j$ over the collinear two-body phase space, which reads
in $d=4-2\epsilon$ dimensions 
\bea\label{eq: nlo-1}
J_{bare}^{(1)}&=& \frac{1}{4}  \frac{1}{(2\pi)^{d-1}} \frac{2\pi^{1-\epsilon} }{\Gamma(1-\epsilon)}
\int \mathrm{d} z \mathrm{d} s_{ij} s_{ij}^{-\epsilon} (z {\bar z})^{1-\epsilon} \nn \\
&&\hspace{-8.ex} \times  \frac{8\pi \alpha_s Z_{\alpha}\,\mu^{2\epsilon} e^{\gamma_E \epsilon}
}{ (4\pi)^\epsilon s_{ij} } 
C_F \left[ \frac{1+ {\bar z}^2}{z} - \epsilon  z \right]  \theta(R^2 -  \Delta R_{ij}^2 )\,,
\eea
where $Z_\alpha$ is the $\alpha_s$ renormalization factor, which at NLO is
\bea
Z_\alpha = 1- \frac{\alpha_s}{4\pi \epsilon } \beta_0 
= 1- \frac{\alpha_s}{4\pi \epsilon} \left( \frac{11}{3} C_A - \frac{4}{3} T_F N_F \right) \,,
\eea
$s_{ij} = 2 p_i \cdot p_j$ and $z$ is the momentum fraction carried by the gluon $g_j$ and $ {\bar z} = 1-z$. The first line of Eq.~(\ref{eq: nlo-1}) is the phase space measure in the collinear limit and we note that the matrix element in second line of Eq.~(\ref{eq: nlo-1}) is nothing but the leading order (LO) splitting function $\frac{8\pi\alpha_s\mu^{2\epsilon}}{s_{ij}}P_{q\to qg}$. The $\theta$-function originates from the requirements of the jet algorithm to cluster two partons into one single jet with radius $R$. Here 
\bea
\Delta R_{ij}^2 = \Delta \eta_{ij}^2 + \Delta\phi_{ij}^2 
\approx \frac{2 p_i \cdot p_j}{ p_{i,T} p_{j,T}} = \frac{s_{ij}}{z{\bar z} p_T^2} \,,
\eea
where $\Delta \eta_{ij}$ and $\Delta \phi_{ij}$ are the rapidity and azimuthal angle differences between parton $i$ and $j$, respectively. 
$p_T$ is the jet transverse momentum and $p_{i,T}$ and $p_{j,T}$ are the transverse momenta for partons $i$ and $j$. The narrow jet approximation $R \ll 1$ has been used.

If we introduce the variables
\bea
x_1 \equiv 
\tilde{s}_{ij} = \frac{s_{ij}}{z {\bar z} (p_T R)^2} \le 1\,, \quad 
x_2 \equiv z \le 1 \,,
\eea
we can write 
\bea
J_{bare}^{(1)} 
&=&  e^{2\epsilon L}    \frac{  \alpha_s   C_F }{2\pi} \frac{Z_\alpha\,e^{\gamma_E \epsilon} }{\Gamma(1-\epsilon)}
\int_0^1  \mathrm{d} x_1 \mathrm{d} x_2 x_1^{-1-\epsilon} x_2^{-1-2\epsilon}   \nn \\
&\times&   (1-x_2)^{-2\epsilon}
\, \left[ 1+(1-x_2)^2 - \epsilon \, x_2^2 \right] \,,
\eea
where $L = \log \frac{\mu}{p_T R}$. All the singularities are given by $x_i
\to 0$ and have been isolated in the factor $x_1^{-1-\epsilon} x_2^{-1-2\epsilon}$ 
so that the remaining terms in the second line of the above equation are finite as $x_i \to 0$.

With this manipulation, to evaluate the integral, 
we can first expand $x_i^{-1-a_i \epsilon}$ using the Laurent expansion
\bea\label{eq:LaurentExpand}
x_i^{-1-a_i \epsilon }
= - \frac{1}{a_i \epsilon} \delta(x_i) + \sum_{n=0} \frac{(-a_i \epsilon )^n}{n!} \left[ \frac{\log^n x_i}{x_i} \right]_+ \,, \quad
\eea
and then perform the integration either numerically or analytically to find the coefficients of the $\epsilon$-poles and the finite terms. This reproduces the well-known NLO bare quark-jet function 
\bea\label{eq:nlojet}
J_{bare}^{(1)} &=&    e^{2\epsilon L}  Z_\alpha \, \frac{\alpha_s}{2\pi} C_F 
\left(
\frac{1}{\epsilon^2} + \frac{3}{2\epsilon} + \frac{13}{2} - \frac{3\pi^2}{4} 
\right. \nn \\
&& + \left[ 26 - \frac{9\pi^2}{8} - \frac{49}{3}\zeta_3
\right]\epsilon 
\nn \\
&& \left.
+ \left[
104- \frac{39}{8}\pi^2 - \frac{49}{2}\zeta_3 - \frac{11}{32}\pi^4
\right]\epsilon^2
\right) \,. \quad
\eea
We note that at NLO, all $k_T$-type jet algorithms give the same results. 

Although we present here only the $x_i$-integrated jet function, our approach
is actually fully exclusive in $x_i$. There are no difficulties to generate
distributions differential in $x_i$, as can be seen from 
Eq.~(\ref{eq:LaurentExpand}), especially in the angular variable $x_1 = \tilde{s}_{ij}$. 
The implications of the possibility to provide exclusive results, especially their role in resumming non-global logarithms, will be discussed later.

A similar strategy will be used to compute the NNLO jet functions but the calculation is more involved, as we will explained in detail in the rest of the work.

\section{NNLO}
\subsection{Real-virtual contribution} 
Now we proceed to calculate the jet function at NNLO. We start with the real-virtual corrections. 
The phase space of the real-virtual corrections is identical to the NLO calculation. The matrix element for the jet function is given by the one-loop correction to the splitting kernel, which is well {documented in~\cite{Kosower:1999xi,Kosower:1999rx,Bern:1999ry,Sborlini:2013jba,Ritzmann:2014mka}.} The calculation for the real-virtual correction is straightforward and gives
\bea\label{eq:nnlojetrv}
J^{(2)}_{rv}
&=& \frac{\alpha_s^2  e^{4\epsilon L} }{(2\pi)^2}  
C_F
\Big(
C_F {\cal K}^{rv}_{C_F} + C_A {\cal K}^{rv}_{C_A} \Big) \,,
\eea
with
\bea
\label{eq:nnlojetrvcf}
{\cal K}^{rv}_{C_F} &=& 
\left( 
-\frac{5}{4}+ \frac{\pi^2}{3}
\right) \frac{1}{\epsilon^2} 
+ \left( 
-\frac{31}{2}+ \frac{\pi^2}{2} + 22 \zeta_3
\right) \frac{1}{\epsilon} \nn \\
&&   
- \frac{575}{4} + \frac{137}{24}\pi^2 + 33\zeta_3 + \frac{10}{9}\pi^4
\, ,
\eea
and
\bea
\label{eq:nnlojetrvca}
{\cal K}^{rv}_{C_A} &=& 
-\frac{1}{4 \epsilon^4}
-\frac{3}{4 \epsilon^3}
+ \left( -5 + \frac{11\pi^2}{24} \right) \frac{1}{\epsilon^2} \nn \\
&&
+\left(- \frac{63}{2} + \frac{13\pi^2}{8} + \frac{26}{3}\zeta_3 \right) \frac{1}{\epsilon}
\nn \\
&& 
- \frac{781}{4}+11\pi^2+\frac{85}{2}\zeta_3 - \frac{67}{1440}\pi^4
\,. \qquad
\eea

\subsection{Real-real contribution}
\subsubsection{Matrix elements}
We turn to the calculation of the real-real corrections. The matrix element involved is nothing but the tree level $a\to ijk$ splitting kernel,
\bea
|{\cal M}|^2 = \left(
\frac{\mu^2 e^{\gamma_E}}{4\pi} \right)^{2\epsilon} \, 
\frac{ 64 \pi^2 \alpha_s^2 }{s^2_{ijk}} P_{a\to ijk}(z_i,z_j,z_k) \,,
\eea
where $s_{ijk} = s_{ij} + s_{ik} + s_{jk} $ and $z_i$ is the momentum fraction
carried by parton $i$, with $z_i + z_j + z_k = 1$. The explicit form of the
splitting function $P_{a\to ijk}(z_i,z_j,z_k)$ can be found in~\cite{Catani:1999ss} (see also \cite{Campbell:1997hg}). 
For the quark-jet we need
\bea
P_{{\bar q}_1' q_2' q_3} \,, \quad 
P_{{\bar q}_1q_2q_3}^{({\rm id})}\,, \quad
P_{g_1g_2q_3}^{({\rm ab})}\,, \quad \text{and} \quad
P_{g_1g_2q_3}^{(\rm nab)} \,,
\eea
and explicit expressions are listed in the appendix~\ref{app:A}.

The matrix element $|{\cal M}|^2$ contains structures
\bea
|{\cal M}|^2 \supset \frac{1}{s_{ijk}^\alpha s_{ij}^\beta} \,, 
\frac{1}{s_{ij} s_{jk}} \,,   
\text{permutations in $i$, $j$, $k$} \,, \quad
\eea
which will develop singularities in $\epsilon$ if one or more of the $s_{ij}$'s vanish upon integration over the phase space. To extract the real-real corrections, our strategy is to write the phase space integration as
\bea\label{eq:psform}
\int d \Phi_3 |{\cal M}|^2 = \int_0^1 \prod_{i}  \mathrm{d}x_i \, 
x_i^{-1-a_i \epsilon} \,
F(\{x_i\},\epsilon) \,, 
\eea
where $F$ is regular as $x_i \to 0$. Hence we can first perform the Laurent
expansion in $x_i^{-1-a_i \epsilon}$ to isolate the $\epsilon$-poles and
compute their coefficients and the remaining constant terms either numerically
or analytically. We emphasize again that Eq.~(\ref{eq:psform}) allows not only
for the integrated jet function but also the differential distributions in 
$x_i$'s. To arrive at the expression in Eq.~(\ref{eq:psform}), we need a proper
parameterization of the phase space.

\subsubsection{Phase Space Parameterization}
In the collinear limit, the three-body phase space can be shown to take the form~\cite{GehrmannDeRidder:1997gf,Ritzmann:2014mka}
\bea
d \Phi_3  &=&  4
\frac{  ds_{ij} ds_{ik} ds_{jk} d z_i dz_j }{(4\pi)^{5-2\epsilon} \Gamma(1-2\epsilon) } \,
 \Delta^{-\frac{1}{2} - \epsilon}   
\,, 
\eea
where the Gram determinant is given by
\bea
\Delta =  4 z_i z_j s_{ik}s_{jk} - (z_k s_{ij}-z_i s_{jk}- z_js_{ik})^2 > 0\,,
\eea
with $z_k = 1- z_i - z_j$. 

Now we introduce the angular variable between two arbitrary partons $a$ and $b$
\bea
\tilde{s}_{ab} = \frac{1}{(p_T R)^2} \frac{s_{ab}}{z_a z_b}
\eea
and further make a variable transformation using one of the $\tilde{s}_{ab}$
\bea\label{eq:var-trans1}
\tilde{s}_{ik}   
= (\sqrt{\tilde{s}_{jk}  } -  \sqrt{ \tilde{s}_{ij}  } )^2 + 
4  \sqrt{   \tilde{s}_{ij} \tilde{s}_{jk}    }t  \,, 
\eea
with $t\in [0,1]$ to find
\bea
d \Phi_3 & = & 
(2p_T R)^{4-4\epsilon} 
\frac{   d\tilde{s}_{ij} d\tilde{s}_{jk}  dt   d z_i dz_j  }{2(4\pi)^{5-2\epsilon} \Gamma(1-2\epsilon) }
   (z_iz_jz_k)^{1 - 2\epsilon}       \nn\\
&& \times 
  \left(   \tilde{s}_{ij} \tilde{s}_{jk}  \right)^{-\epsilon }  
t^{-\frac{1}{2} - \epsilon}   
\left(
   1-t
\right)^{-\frac{1}{2} - \epsilon}   
\,.
\eea
If $\tilde{s}_{ik}$ happens to be in the denominator of the matrix element, the variable change in Eq.~(\ref{eq:var-trans1}) will develop a line singularity, which can be mapped to $\tilde{s}_{ij} = \tilde{s}_{jk}$ by a non-linear transformation~\cite{Anastasiou:2003gr,Boughezal:2015eha} 
\bea
x_5' = \frac{ (\sqrt{\tilde{s}_{ij}  } -  \sqrt{ \tilde{s}_{jk}  } )^2 (1-t) }{\tilde{s}_{ik}} \,,
\eea
and thus 
\bea
\tilde{s}_{ik} = (\tilde{s}_{ij} - \tilde{s}_{jk})^2 
\left( ( \sqrt{\tilde{s}_{ij}}  - \sqrt{ \tilde{s}_{jk} } )^2 + 4 \sqrt{ \tilde{s}_{ij} \tilde{s}_{jk}}  \,   x_5' \right)^{-1} \,, 
\nn \\
\eea
which gives
\bea
d \Phi_3  &=&      (2p_T R)^{4-4\epsilon}  \, 
\frac{\pi\,   d\tilde{s}_{ij} d\tilde{s}_{jk}  d z_i dz_j  dx_5   }{2 (4\pi)^{5-2\epsilon} \Gamma(1-2\epsilon) }
   (z_iz_jz_k)^{1 - 2\epsilon}       \nn \\
&\times&  
  \left(   \tilde{s}_{ij} \tilde{s}_{jk}  \right)^{-\epsilon }  \,
 {x_5'}^{-\epsilon } (1-x_5')^{-\epsilon }  \,
|\tilde{s}_{ij}-\tilde{s}_{jk}|^{1-2\epsilon} \nn 
\\
&\times & 
\left(
 ( \sqrt{\tilde{s}_{ij}}  - \sqrt{ \tilde{s}_{jk} } )^2 + 4 \sqrt{ \tilde{s}_{ij} \tilde{s}_{jk}}  \,   x_5'
  \right)^{-1+2\epsilon}
\,.
\eea
Here we have introduced the variable $x_5$ through 
\bea
x_5' = \sin^2\left( \frac{\pi}{2} x_5 \right)\,.
\eea
Without loss of generality, we can assume $z_i \le z_j$ and $\tilde{s}_{ij} \le \tilde{s}_{jk}$, and in this case we can then introduce~\footnote{It is understood that if $x_i$ is not in $[0,1]$ but $[0,a]$ instead, we can do $x_i \to a x_i$ to map it onto $[0,1]$} 
\bea
\tilde{s}_{jk} = x_1\,, \quad
\tilde{s}_{ij} = x_1  x_2 \,, \quad
z_j = x_3 \,, \quad
z_i = x_3 x_4 \,, \quad
\eea
to reach the phase space parameterization
\bea\label{eq:psparam}
d \Phi_3  &=&     (2p_T R)^{-4\epsilon}   
\frac{ \pi \,   d x_1 d x_2  dx_3   d x_4 dx_5   }{2(4\pi)^{5-2\epsilon} \Gamma(1-2\epsilon) }    \times z_k \nn \\
& \times  & \left[  z_k^{2}\, 
  \, 
x'_5  (1- x'_5) \,  
\left(
 ( \sqrt{    x_2  }  - 1 )^2 + 4 \sqrt{   x_2 }  \,   x_5'
  \right)^{-2}  \right]^{-\epsilon } \nn \\
&\times &  x_1^{-1-2\epsilon}\,  x_2^{-1-\epsilon }  \, 
  (1-x_2)^{-1-2\epsilon}   \, 
  x_3^{-1-4\epsilon} \, 
    \,     x_4^{-1 - 2\epsilon}   \, 
    \nn \\
  & \times &
\left[  x_1^2 x_2 (1-x_2)^2 x_3^4 x_4^2 
  ( (1-\sqrt{x_2})^2+4\sqrt{x_2}x_5' )^{-1}  \right]
\,.
\nn \\
\eea
Other orders can be obtained in the same manner. In the third line of the phase space parameterization, the Laurent expansion 
in Eq.~(\ref{eq:LaurentExpand}) can be performed to extract the $\epsilon$-poles and the finite contributions. 
On the other hand, the last line of Eq.~(\ref{eq:psparam}), when multiplied
with the matrix element, will be free of any singularities as $x_i \to 0$ or
$x_2 \to 1$. It is suitable for a numerical evaluation, as long as appropriate
subtraction terms are constructed as will be explained in the following section.

\subsubsection{Jet algorithm and subtraction terms}
In this work, we are always interested in $k_T$-type jet algorithms, which involves the comparison of the metrics
\bea
\rho_{ij} 
= \min\left[ p_{T,i}^{-2\alpha}, p_{T,j}^{-2\alpha }\right]
\frac{\Delta R_{ij}^2}{R^2}
\,,\quad
\rho_i  = p_{T,i}^{-2\alpha} \,,
\eea
for some parameter $\alpha$ and $\alpha=1$ defines the anti-$k_T$ jet algorithm.   

In the small jet radius limit, or in the collinear limit, we can approximate $p_{T,i} = z_i \, p_{T}$ up to power corrections. 
Furthermore, if we demand all three partons to be grouped into the same jet,
we will require in the last stage of the clustering, 
\bea\label{eq:cluster-all}
\Delta R_{\tilde{ij},k} &=& 
\Delta \eta_{\tilde{ij},k}^2 +
 \Delta \phi_{\tilde{ij},k}^2   \nn \\
&\approx& 2 ( \cosh \Delta \eta_{\tilde{ij},k} - \cos \Delta \phi_{\tilde{ij},k} )  \le R^2 \,,
\eea
where we have assumed that the partons $i$ and $j$ are clustered first into the pseudo-particle $\tilde{ij}$, 
which then clusters with parton $k$ to form the final jet.

Within the E-scheme recombination, the four-momentum of the pseudo particle $\tilde{ij}$ is given by $p^\mu_{\tilde{ij}} = p^\mu_i + p^\mu_j$ and thus
\bea
&& 2 p_{\tilde{ij}} \cdot p_k 
=  2 p_{k,T} (m_{\tilde{ij},T}  \cosh \Delta\eta_{\tilde{ij},k} - 
 p_{\tilde{ij},T}   \cos \Delta\phi_{\tilde{ij},k}  
) \, \nn \\
&=& 
2 p_{k,T} \left(\sqrt{ p^2_{\tilde{ij},T}   + s_{ij}}  \cosh \Delta\eta_{\tilde{ij},k} - 
 p_{\tilde{ij},T}   \cos \Delta\phi_{\tilde{ij},k}  
\right) \nn \\
&=&2 p_{k,T} p_{\tilde{ij},T} (  \cosh \Delta\eta_{\tilde{ij},k} - 
 \cos \Delta\phi_{\tilde{ij},k}  
) 
+  p_{k,T} \frac{s_{ij}}{p_{\tilde{ij},T}}
\nn \\
& &\qquad\qquad\qquad\qquad\qquad\qquad
\times   \left(
1 + {\cal O} \left( s_{ij}/p^2_{{\tilde{ij}},T} \right)
\right)
\nn \\
&\approx &2 p_T^2 (z_i+z_j) z_k (  \cosh \Delta\eta_{\tilde{ij},k} - 
 \cos \Delta\phi_{\tilde{ij},k}  
)  \nn \\
&& + \frac{z_k z_i z_j }{(z_i+z_j)} \frac{s_{ij}}{z_i z_j p_T^2 R^2} p_T^2 R^2
 \,. 
\eea
From this, it is easy to see that the requirement in Eq.~(\ref{eq:cluster-all}) of all three partons being clustered into one jet can be written as
\bea\label{eq:cluster-c}
c_{ijk} \equiv 
\left( z_i \tilde{s}_{ik} + z_j \tilde{s}_{jk}  \le z_i + z_j 
+ \frac{z_iz_j}{z_i+z_j} \tilde{s}_{ij} \right)
\,,
\eea
valid in the small-$R$ limit. 
Here again, $\tilde{s}_{ij} \equiv \frac{s_{ij} }{z_i z_j} \frac{1}{p_{T}^2 R^2 }$.

In the earlier history of the clustering, in order to have $i$ and $j$ to be clustered first, we need  
\bea
\rho_{ij} < \min\left[ 
\rho_{ik} \,, \rho_{jk} \,,
\rho_i\,,\rho_j\,,\rho_k
\right] \,,
\eea
as required by the jet algorithm. 
This maps to the conditions for the following cases:
\begin{itemize}
\item case I: \\
$\tilde{s}_{ij}
<  z_j^{ 2\alpha} 
z_k^{-2\alpha}  \min [
\tilde{s}_{ik},
\tilde{s}_{jk}, 1
]
$, for $z_i \le z_j \le z_k$;

\item case II: \\ $\tilde{s}_{ij}
<   \min [  
z_k^{-2\alpha}  z_j^{2\alpha}     \tilde{s}_{ik},
  \tilde{s}_{jk},
 1  ]$, for $z_i \le z_k \le z_j$;
 
\item case III: \\ $\tilde{s}_{ij}
<  \min [  z_j^{2\alpha} 
 z_i^{-2\alpha} 
\tilde{s}_{ik},
\tilde{s}_{jk},
1]$, for $z_k \le z_i \le z_j$. 
\end{itemize}
Other cases can be obtained by switching the order between $i$ and $j$.

The jet clustering conditions can be expressed as constraints in the phase space integrals. For case I, we have
\bea
&&d \Phi^{\text{I}}_3
= 
d \Phi_3\,   \theta\left( \tilde{s}_{ij} < \left( \frac{z_j}{z_k}\right)^{2\alpha} \tilde{s}_{jk} \right) 
  \theta\left(
   \tilde{s}_{ij} < \left( \frac{z_j}{z_k} \right)^{2\alpha}  \tilde{s}_{ik} 
   \right) \,
  \nn \\
&&  \times {\theta\left(
\tilde{s}_{ij} < \left( \frac{z_j}{z_k} \right)^{2\alpha}   
\right) }
  \theta\left( c_{ijk} \right) \theta(z_i \le z_j \le z_k) \,,
\eea
where $d\Phi_3$ is the three-body phase space given in the previous
section and $c_{ijk}$ is given in Eq.~(\ref{eq:cluster-c}). We note that the $\theta$-functions in the first line due to the
clustering condition are dangerous, in the sense that they will change the
pre-determined fractional power of $\epsilon$ in the exponent in Eq.~(\ref{eq:psparam}), associated with the $z_j \to 0$ poles and therefore invalidate the Laurent expansion in Eq.~(\ref{eq:LaurentExpand}). To understand the issue, we consider a simple example by comparing the integrals 
\bea
I_1 = \int_0^1   \mathrm{d}x_3 \, \mathrm{d} x_2 \, x_2^{-1-a_2 \epsilon} 
x_3^{-1-a_3 \epsilon} \,, 
\eea
and 
\bea
I_2 = \int_0^1  \mathrm{d}x_3 \mathrm{d} x_2 \, x_2^{-1-a_2 \epsilon} 
x_3^{-1-a_3 \epsilon}  \theta( x_3^\alpha -  x_2 )\,.
\eea
Apparently we can safely Laurent-expand the integrand $x_2^{-1-a_2 \epsilon} 
x_3^{-1-a_3 \epsilon}$ in $I_1$ to perform the integration. 
This naive expansion is not allowed in $I_2$, since the correct power of $\epsilon$ for $x_3$ is not $- a_3 \epsilon$
but $-(a_3 + a_2 \alpha) \epsilon $, which can be easily seen by integrating out $x_2$. To avoid this complication, we introduce a subtraction term, to remove all possible the $z_j \to 0$ singularities by taking the $z_j \to 0$ limit in the clustering condition to find
\bea
d \Phi^{\text{I}}_{3,sub.}
&=& 
 d \Phi_3 \,  \theta\left( \tilde{s}_{ij} < z_j^{2\alpha} \tilde{s}_{jk} \right) 
 \,
  \theta\left(  \tilde{s}_{jk}   \le 1    \right)   
  \theta\left(  \tilde{s}_{ik}   \le 1    \right)   \nn \\
&& \times \theta(z_i \le z_j \le z_k)  \,,
\eea
where we have used the fact that as $z_j \to 0$, $\tilde{s}_{ij} \to 0$ as forced by the $\theta$-functions, and therefore
$\tilde{s}_{jk} \to \tilde{s}_{ik}$. With the subtraction, the integral
$(d\Phi_3^I - d \Phi^{I}_{3,sub.})|{\cal M}|^2$ will be free of the $z_j \to
0$ poles. The term $d \Phi^{I}_{3,sub.} \, |{\cal M}|^2 $ alone still contains the $z_j \to 0$ singularity and therefore the potential $\epsilon$ fractional power modification due to $\theta\left( \tilde{s}_{ij} < z_j^{2\alpha} \tilde{s}_{jk} \right)$. But we will see later that when combined with other subtraction terms introduced soon, all such dangerous $\theta$-functions go away.

We have, for case II and case III, the phase space integrals
\bea
 d \Phi^{\text{II}}_3 
&=& d \Phi_3 
 \left[1- \theta\left(   \tilde{s}_{ij}
  <   \left( \frac{z_j}{z_k} \right)^{2\alpha } \tilde{s}_{ik}\right)    \right] 
  \theta( \tilde{s}_{ik} <1 )
  \nn \\
&& \times
\theta( \tilde{s}_{ik} \le \tilde{s}_{jk}) 
\theta\left( 
c_{ikj}
\right) 
\theta(z_i \le z_j \le z_k) \,,
\eea
and 
\bea
d \Phi_3^{\text{III}}
&= & d \Phi_3 
\times \left[1-  \theta\left(   \tilde{s}_{ij} 
 <   \left( \frac{z_j}{z_k} \right)^{2\alpha } \tilde{s}_{jk} \right)  \right]
\theta( \tilde{s}_{jk} <1 )
 \nn \\
&&\times 
\theta( \tilde{s}_{jk} < \tilde{s}_{ik})
\theta\left(
c_{jki}  \right)
 \theta(z_i \le z_j \le z_k)\,,
\eea
respectively. Here for later use, we have relabeled subscripts $i$, $j$ and
$k$ to stick to the case that $z_i \le z_j \le z_k $, but with $i$, $k$
clustered first in case II and $j$ and $k$ first in case III. This is always
allowed since eventually we sum over all possible orders of $z_i$'s and all possible clustering histories. Following the same logic as in case I, we construct the subtraction terms for case II
\bea
d \Phi^{\text{II}}_{3,sub.} \,=\, d \Phi_3 
\left[1 - \theta\left(   \tilde{s}_{ij}
  <   z_j^{2\alpha } \tilde{s}_{jk}\right)    \right] 
\theta( \tilde{s}_{ik} \le \tilde{s}_{jk}) \nn \\
\times\, 
\theta( \tilde{s}_{ik} <1 )
 \theta\left(    \tilde{s}_{jk} < 1   \right) \theta(z_i < z_j<z_k) \,, 
\qquad 
\eea
and case III
\bea
d\Phi_{3,sub.}^{\text{III}} \,=\, d \Phi_3 
 \left[1-  \theta\left(   \tilde{s}_{ij} 
 <   z_j^{2\alpha } \tilde{s}_{jk} \right)  \right]  
\theta( \tilde{s}_{jk} < \tilde{s}_{ik}) \nn \\
\times\, 
\theta( \tilde{s}_{jk} <1 )
\theta\left(  \tilde{s}_{ik} < 1 \right)\times \theta(z_i < z_j < z_k ) \,, 
\qquad 
\eea
respectively, to remove the $z_j \to 0$ singularities. 

The phase space integral with the jet algorithm implemented reads
\bea
&& d\Phi_{\rm alg.} |{\cal M}|^2 = \sum_{i=\text{I}}^{\text{III}} \left(
d \Phi_3^{i} -  d \Phi_{3,sub.}^{i}  \right)  |{\cal M}|^2
+\, d \Phi_{3,sub.}^{i}  |{\cal M}|^2
\nn \\
& &\hspace{5.ex} 
+\left(\text{all $z_i$, $z_j$, $z_k$ orders}
\right) \,, 
\eea 
and we note that by construction, it is easy to check that
\bea\label{eq:inclusive}
\sum_{i={\text I}}^{\text{III}} d \Phi_{3,sub.}^{i}  
=
d \Phi_3    \theta( \tilde{s}_{jk} <1 )
\theta\left(  \tilde{s}_{ik} < 1 \right) \, \theta(z_i < z_j < z_k )  \,, \nn \\
\eea
where all $\theta$-function constraints involving powers of $z_j^\alpha$ cancel in the sum.
Therefore, the phase space integration does not induce any modifications 
of the fractional powers in $\epsilon$ as $z_j \to 0$. 

However at this stage, the phase space parameterization in Eq.~(\ref{eq:psparam}) is still not sufficient to isolate all singularities within $|{\cal M}|^2$. Particularly in $\sum_{i={\text I}}^{\text{III}} d \Phi_{3,sub.}^{i}$, a further ordering between the variables $x_2$ and $x_4$ is required to fully disentangle the poles generated by 
$s_{ijk} = z_i z_k \tilde{s}_{ik} + z_jz_k \tilde{s}_{jk}+ z_iz_j\tilde{s}_{ij}\to 0$. 
{To avoid such an ordering, we introduce instead a subtraction term in the matrix element 
to regulate the single soft limit, as $x_4 \to 0$, of $|{\cal M}|^2$.
Thus, a natural choice for the subtraction term $|{\cal M}|_{sub.}^2$ is}
\bea
|{\cal M}|_{sub.}^2 = \lim_{x_4 \to 0} |{\cal M}|^2
= T_j \cdot T_k\, \frac{s_{jk}}{s_{ij} s_{ik}} P^{(0)}_{a \to jk} \,,
\eea
which is nothing but the product of the LO eikonal factor and the LO splitting kernel. 

The final phase space integration suitable for a numerical evaluation is thus given by
\bea
 d\Phi_{\rm alg.} |{\cal M}|^2 &=&
  \sum_{i=\text{I}}^{\text{III}}  d \Phi_{3,sub.}^{i}  |{\cal M}|_{sub.}^2 
\nn \\
& &
  + 
  \left(
d \Phi_3^{i} -  d \Phi_{3,sub.}^{i}  \right)  |{\cal M}|^2 \nn \\
& & +\,
 \sum_{i=\text{I}}^{\text{III}}
 d \Phi_{3,sub.}^{i} ( |{\cal M}|^2 -  |{\cal M}|_{sub.}^2)
\nn \\
& &+
\left(\text{all $z_i$, $z_j$, $z_k$ orders}
\right) \,.
\eea 
Similar subtraction counter terms can be constructed for the WTA scheme recombination as well as other jet algorithms such as the Cambridge-Aachen algorithm and thus used to compute the soft-drop groomed jet function. Further applications to obtain the semi-inclusive jet function~\cite{Kang:2016mcy, Dai:2016hzf} at ${\cal O}(\alpha_s^2)$ are also straightforward.

\subsubsection{Results and validations}
With the setups developed in the previous sections, we find the double-real correction to the anti-$k_T$ jet function to be 
\bea
\label{eq:nnlojetrr}
 J_{rr}^{(2)} = \, \frac{\alpha_s^2  e^{4\epsilon L} }{(2\pi)^2}  
C_F
\Big(
C_F {\cal K}^{rr}_{C_F} + C_A {\cal K}^{rr}_{C_A}
+ N_F T_F {\cal K}^{rr}_{N_F T_F} \Big) \,,  \nn \\
\eea
where we find
\bea
\label{eq:nnlojetcf}
{\cal K}^{rr}_{C_F} &=& \frac{1}{2\epsilon^4} + \frac{3}{2\epsilon^2}
- \frac{1.8171(3)  }{\epsilon^2} 
- \frac{ {21.272(3)}  }{\epsilon} 
- {76.42(2)}   \,,  \nn 
%
\\[1ex]
\label{eq:nnlojetca}
{\cal K}^{rr}_{C_A} &=&  \frac{1}{4\epsilon^4} + 
\frac{1.20833 
}{\epsilon^3} 
 + \frac{1.5484(2)  }{\epsilon^2} 
 \nn \\
 &&
 - \frac{{7.941(4)}  }{\epsilon}
 -{ 75.425(2) }
\,,
\eea
and
\bea\label{eq:nnlojetnftf}
{\cal K}^{rr}_{N_F T_F} &=&  
- \frac{1}{6\epsilon^3} - \frac{7}{9 \epsilon^2} 
+ \frac{0.1067(3)  }{\epsilon} 
+ { 17.230(2) } 
\,.
\qquad 
\eea
Here and below we always include in brackets the numerical error on the last digit.

To validate our results, we have performed independent calculations and found full agreement. Also we note that parts of our calculations associated with  $d\Phi_{3,sub.}$ can be done (semi-)analytically and for these parts, we have checked our numerical computation against the (semi-)analytic results to find full agreement. Some of these checks are presented in the appendix~\ref{app:C}. 

\subsubsection{Renormalization group equation}

The renormalization group equation (RGE) for the quark-jet function serves as
strong independent check on the calculation. 
The leading poles up to $\epsilon^{-2}$ of the double-real correction in Eq.~(\ref{eq:nnlojetrr}) 
combined with the real-virtual one in Eq.~(\ref{eq:nnlojetrv}) 
can be predicted by solving the RGE up to the logarithmic term $\alpha_s^2L^2$
and taking the NLO quark-jet function from Eq.~(\ref{eq:nlojet}) as input.
Details are given in the appendix~\ref{app:B}. 

As a result, the RGE predicts the leading three poles normalized to $\alpha_s^2/(2\pi)^2$ 
for the $C_F^2$ term as
\bea
{\cal K}^{rr}_{C_F} \biggr|_{\epsilon^{-4}} &=& 
\frac{1}{2\epsilon^4} \,, 
\\
{\cal K}^{rr}_{C_F} \biggr|_{\epsilon^{-3}} &=& 
\frac{3}{2\epsilon^3} \,, 
\\
\label{calKcfepm2}
\big({\cal K}^{rv}_{C_F} + {\cal K}^{rr}_{C_F}\big) \biggr|_{\epsilon^{-2}} &=& 
\left(\frac{61}{8} - \frac{3\pi^2}{4} \right) \frac{1}{\epsilon^2}
\approx \frac{0.222797}{\epsilon^2} 
\nn \\
&=&\frac{0.2228(3)   }{\epsilon^2} \,, 
\qquad
\eea
which all agree exactly with Eqs.~(\ref{eq:nnlojetrvcf}) and (\ref{eq:nnlojetcf}), 
in the case of the $\epsilon^{-2}$ pole within numerical errors as shown in
the second line in Eq.~(\ref{calKcfepm2}).

For the $C_F N_F T_F$ contribution, solving the RGE we find the leading poles
\bea
{\cal K}^{rr}_{N_F T_F} \biggr|_{\epsilon^{-3}} &=& - \frac{1}{6\epsilon^3}
\,,
\\
{\cal K}^{rr}_{N_F T_F} \biggr|_{\epsilon^{-2}} &=& -\frac{7}{9\epsilon^2} \,,
\eea
which fully agrees with what we obtained in Eq.~(\ref{eq:nnlojetnftf}). 

As for the $C_F C_A$ term, the naive RGE alone can predict the
$\epsilon^{-4}$ and $\epsilon^{-3}$ poles correctly, however the
$\epsilon^{-2}$ term is affected by the non-global contribution, for which we find
that an additional $-\frac{\pi^2}{12 \epsilon^2}$ correction is needed to produce
the correct pole and the logarithm at $\alpha_s^2L^2$. 
The same situation has been encountered in the Sterman-Weinberg cone jet in~\cite{Becher:2015hka} 
and for the hemisphere mass distribution in \cite{Schwartz:2014wha}.

Once every piece is taken into account, the poles predicted for the $C_F C_A$ contribution are
\bea
\big({\cal K}^{rv}_{C_A} + {\cal K}^{rr}_{C_A}\big) \biggr|_{\epsilon^{-4}}
&=& 0\, ,
\\
\big({\cal K}^{rv}_{C_A} + {\cal K}^{rr}_{C_A}\big) \biggr|_{\epsilon^{-3}}
&=&
\frac{11}{24 \epsilon^3}
\approx \frac{0.458333}{\epsilon^3}
\nn \\
&=&
\frac{0.45833(5) }{ \epsilon^3} \,, \quad
 \\
\label{calKcaepm2}
\big({\cal K}^{rv}_{C_A} + {\cal K}^{rr}_{C_A}\big) \biggr|_{\epsilon^{-2}}
&=&
\left( \frac{83}{36} - \frac{\pi^2}{8} \right) \frac{1}{\epsilon^2} 
\approx \frac{1.07186}{\epsilon^2} 
\nn \\
&=& 
\frac{1.0720(2)    }{\epsilon^2} 
\,, 
\eea
where the agreement is within numerical errors.
For the $\epsilon^{-2}$ term it is at ${\cal O}(1\mbox{\textperthousand})$ 
as shown in the second line in Eq.~(\ref{calKcaepm2}).

\section{Two-loop jet function and discussions}
Now we present the anti-$k_T$ bare quark-jet function at two loops 
\bea
J_{bare} &=& 1 + J_{bare}^{(1)} + \frac{\alpha_s^2e^{4 \epsilon L}}{4\pi^2} 
C_F
\Big(
C_F {\cal J}_{C_F} \nn \\
&& \hspace{9.ex} + C_A {\cal J}_{C_A}
+ N_F T_F {\cal J}_{N_F T_F} \Big) \,,
\eea
where $J^{(1)}_{bare}$ has been given in Eq.~(\ref{eq:nlojet}).
The new two-loop result reads 
\bea
{\cal J}_{C_F} &=& \frac{1}{2\epsilon^4} + \frac{3}{2\epsilon^3} 
+ \frac{1}{\epsilon^2} \left( \frac{61}{8}  - \frac{3 \pi^2}{4} \right) 
\nn \\
&&
- \frac{ {5.392(3)}}{\epsilon} - { 15.93(2)}  \,, 
\\
%
{\cal J}_{C_A} &=& \frac{11}{24\epsilon^3} 
+ \frac{1}{\epsilon^2} \left( \frac{83}{36}  - \frac{ \pi^2}{8} \right) 
- \frac{ {12.985(4)}}{\epsilon}  - { 115.55(2)}  \,, \nn\\
\eea
and 
\bea
{\cal J}_{N_FT_F} =- \frac{1}{6\epsilon^3} 
- \frac{7}{9\epsilon^2} 
+ \frac{0.1067(3)}{\epsilon} + { 17.230(2) } \,,
\eea
with additional ${\cal O}(\alpha_s^2)$ contributions coming from the $\alpha_s$ renormalization of $J_{bare}^{(1)}$ in Eq.~(\ref{eq:nlojet}). 
The renormalized jet function $J$ is then found to be
\bea
\label{eq:Jren}
J &=& 1+ \frac{\alpha_s}{2\pi} C_F \left( \frac{13}{2} - \frac{3\pi^2 }{4}\right) 
+ \frac{\alpha_s^2}{4\pi^2} \left(
 {-1.78(2) }  C_F^2 
\right.
\nn \\
&& \left.
{-106.87(2)} C_A C_F  + { 14.072(2)} C_F N_F T_F 
\right) \,,  \qquad
\eea
where we have set the scale $\mu = p_T R$. 
The associated numerical errors originate entirely from the NNLO real-real corrections. 

\begin{figure}
\includegraphics[width=\linewidth]{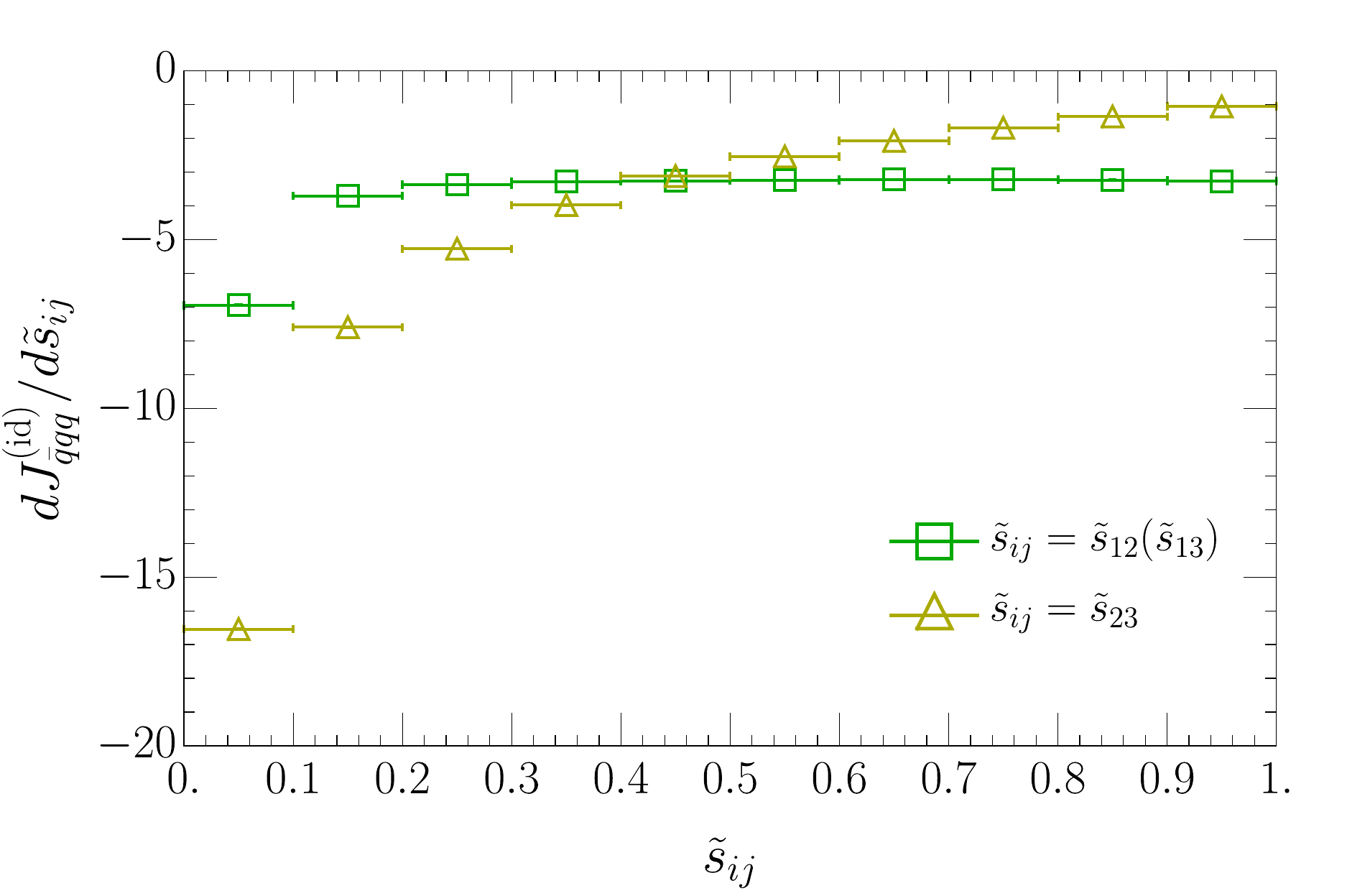} 
\caption{{The distribution $dJ^{({\rm id})}_{{\bar q}qq}/d\tilde{s}_{ij}$ 
for the $\epsilon^0$ part of the jet function $J_{m=3}^{(0)}$ with $P^{({\rm id})}_{{\bar q}_1q_2q_3}$ term only.}}
\label{fig:dist}
\end{figure}

Before concluding we make two comments on our calculation and the result.
First, although we present here the jet function $J$ in
the simplified factorization in Eq.~(\ref{eq:factorizationnll}), there is no
difficulty to provide also the angular dependent jet function $J_{m}$ in
Eq.~(\ref{eq:factorization}) with explicit dependence on the collinear parton
multiplicity $m$ up to $m=3$. 
In fact, the results of $J_{m=2}^{(0)}$, $J_{m=2}^{(1)}$ and $J_{m=3}^{(0)}$ are already encoded in our calculation, 
which can be easily seen from the fully differential nature of the phase space
sector decomposition and the relation that $J = \sum_m \langle J_m \rangle_\Omega $. 
{To demonstrate such feasibility, we plot in
  fig.~\ref{fig:dist}, the distribution in $\tilde{s}_{ij}$ 
  for the $\epsilon^{0}$ part of the jet function $J_{m=3}^{(0)}$ 
  with the $P^{({\rm id})}_{{\bar q}_1q_2q_3}$ contribution only. A detailed study of the angular-dependent jet function will be left for future work.}
The results for the individual $J_m$ will be more suitable for resumming 
NGLs and for improving the logarithmic accuracy of the predictions.

Second, our two-loop result confirms that the factorization theorem in 
Eq.~(\ref{eq:factorizationnll}), as used in many phenomenological studies of jets, TMDs and in spin physics,
is only valid up to the single logarithmic level $({\cal O}\left( e^{\alpha_s^n L^n} \right))$ due to the existence of the NGLs, which is generic for many exclusive jet processes. 
Beyond this order, the factorization in Eq.~(\ref{eq:factorization}) 
or alternative approaches~\cite{Nagy:2020rmk} should be used for predictions. 
Our calculational framework can be useful for realizing the
resummation in Eq.~(\ref{eq:factorization}). 
We leave this for future studies.

\section{Conclusions}
In this work, we have developed a method to calculate the jet functions, which
arise in the factorization of cross sections with exclusive jets of small radius $R$,
with a jet clustering procedure up to NNLO.
Explicit results at NNLO have been presented, for the first time, for the 
quark-jet function with the anti-$k_T$ jet algorithm.
The new two-loop contribution to the anti-$k_T$ quark-jet function provides the 
missing input for the evaluation of the factorization theorem in
Eq.~(\ref{eq:factorization}) to NNLO and for resummation beyond NLL accuracy.
This will substantially improve the precision of predictions for 
observables with jets as hard probes. 
The computational framework is not limited to anti-$k_T$ E-scheme jets 
but readily applicable to many other jet observables up to NNLO, 
such as the semi-inclusive jet function, the WTA jet and the soft-drop groomed
jet radius $R_g$ and the jet momentum fraction $z_g$. 
In addition, the obtained two-loop quark-jet function will serve as an important building block 
when using the factorization in Eq.~(\ref{eq:factorization}) for the development of a colorful $q_T$-subtraction scheme, 
which is applicable to both, calculations at fixed order in perturbation theory up to NNLO and to parton shower simulations. 

In summary, given the broad interest in jets from both the high-energy particle and the nuclear communities, 
we expect the results to have numerous phenomenological applications at
colliders, such as the LHC and a future EIC for the study of 
jet cross sections, TMDs and in addressing the spin structure of the nucleon.

\section*{Acknowledgments} 
We thank Ding-Yu Shao for clarifications on the winner-take-all jet. X.L. is supported by National Natural Science Foundation of China under Grant No.11775023. 
S.M. is supported by Deutsche Forschungsgemeinschaft (DFG) through the
Research Unit FOR 2926, ``Next Generation pQCD for Hadron Structure: Preparing
for the EIC'', project MO 1801/5-1. We thank Kyle Lee, Jürg Haag, Luca Buonocore, Luca Rottoli and Massimiliano Grazzini for pointing out the inappropriate approximation in the early version of the draft. The method proposed in this work are not affected, while the numerics in the concrete example are corrected.

\begin{widetext}

\appendix
\section{The $a\to ijk$ splitting functions}
\label{app:A}
The splitting functions used in our calculation are taken from~\cite{Catani:1999ss}.
For the case in which the quark-anti-quark pair ${\bar q}_1'q_2'$ has a different flavor from the quark $q_3$, 
we use 
\bea
P_{{\bar q}_1' q_2' q_3} = \frac{1}{2} C_F T_F \frac{s_{123}}{s_{12}}
\left[-
\frac{t_{12,3}^2}{s_{12}s_{123}}
+ \frac{4z_3+(z_1-z_2)^2}{z_1+z_2}
+(1-2\epsilon) \left( z_1+z_2 - \frac{s_{12}}{s_{123}} \right)
\right] \,,
\eea
with the abbreviation 
\bea
t_{ij,k} \equiv 2 \frac{z_i s_{jk} - z_j s_{ik}}{z_i+z_j}
+ \frac{z_i-z_j}{z_i+z_j} s_{ij} \,.
\eea
For the final state with identical quark flavors ${\bar q}_1 q_2 q_3$, we need
\bea
P_{{\bar q}_1 q_2 q_3} =  \Big[ P_{{\bar q}_1' q_2' q_3} + (2\leftrightarrow 3) \Big]
+ P_{{\bar q}_1 q_2 q_3}^{(\rm id)} \,,
\eea
where $P_{{\bar q}_1 q_2 q_3}^{(\rm id)} $ accounts for the interference term 
\bea
P_{{\bar q}_1 q_2 q_3}^{(\rm id)}  &=&
C_F \left( C_F - \frac{C_A}{2} \right) \, \Bigg\{
(1-\epsilon) \left( \frac{2s_{23}}{s_{12}} - \epsilon \right)
+ \frac{s_{123}}{s_{12}}\left[
\frac{1+z_1^2}{1-z_2} - \frac{2z_2}{1-z_3} - \epsilon\left(
\frac{(1-z_3)^2}{1-z_2} + 1+z_1 - \frac{2z_2}{1-z_3}
\right) 
\right.
\nn \\
&& 
\left.
- \epsilon^2(1-z_3) 
\right] 
-\frac{s_{123}^2}{s_{12}s_{13}} \frac{z_1}{2}
\left[
\frac{1+z_1^2}{(1-z_2)(1-z_3)}
- \epsilon \left( 1+2 \frac{1-z_2}{1-z_3} \right) - \epsilon^2 \right]  \Bigg\}
+(2\leftrightarrow 3) \,. 
\eea

The splitting function for the quark decay into $g_1g_2q_3$ is given by
\bea
P_{g_1g_2q_3} = C_F^2 P_{g_1g_2q_3}^{({\rm ab})}
+ C_F C_A P_{g_1g_2q_3}^{({\rm nab})} \,,
\eea
where the Abelian contribution is 
\bea
P_{g_1g_2q_3}^{({\rm ab})} &=& \Bigg\{
\frac{s_{123}^2}{2s_{13}s_{23}} z_3
\left[
\frac{1+z_3^2}{z_1z_2}
- \epsilon \frac{z_1^2+z_2^2}{z_1z_2}
-\epsilon(1+\epsilon) 
\right] 
+(1-\epsilon) \left[
\epsilon - (1-\epsilon) \frac{s_{23}}{s_{13}}
\right]
\nn \\
&&+  \frac{s_{123}}{s_{13}} \left[
\frac{z_3(1-z_1)+(1-z_2)^3}{z_1z_2}
+ \epsilon^2(1+z_3) - \epsilon (z_1^2+z_1z_2+z_2^2) \frac{1-z_2}{z_1z_2}
\right] \Bigg\} + (1\leftrightarrow 2) \,, 
\eea
and the non-Abelian one reads
\bea
P_{g_1g_2q_3}^{({\rm nab})}
&=& \Bigg\{ (1-\epsilon) \left( 
\frac{t_{12,3}^2}{4s_{12}^2} + \frac{1}{4} - \frac{\epsilon}{2}
\right)
+ \frac{s_{123}^2}{2s_{12}s_{13}}
\left[
\frac{(1-z_3)^2(1-\epsilon)+2z_3}{z_2}
+ \frac{z_2^2(1-\epsilon)+2(1-z_2)}{1-z_3}
\right] \nn \\
&& - \frac{s_{123}^2}{4s_{13}s_{23}}\, z_3 \left[
\frac{(1-z_3)^2(1-\epsilon)+2z_3}{z_1z_2}
+ \epsilon(1-\epsilon) 
\right] \nn \\
&& + \frac{s_{123}}{2s_{12}} \left[
(1-\epsilon) \frac{z_1(2-2z_1+z_1^2)-z_2(6-6z_2+z_2^2)}{z_2(1-z_3)}
+ 2 \epsilon \frac{z_3(z_1-2z_2)-z_2}{z_2(1-z_3)}
\right] \nn \\
&&+ \frac{s_{123}}{2s_{13}} \left[
(1-\epsilon) \frac{(1-z_2)^3+z_3^2-z_2}{z_2(1-z_3)}
- \epsilon \left(
\frac{2(1-z_2)(z_2-z_3)}{z_2(1-z_3)} - z_1 + z_2
\right) \right. \nn \\
&& \hspace{8.ex} \left. - \frac{z_3(1-z_1)+(1-z_2)^3}{z_1z_2}
+ \epsilon (1-z_2) \left( \frac{z_1^2+z_2^2}{z_1z_2} - \epsilon \right)
\right]
\Bigg\} +(1\leftrightarrow 2) 
\,.  
\eea

\section{The leading $\epsilon$-poles and large logarithms from the RGE}
\label{app:B}
Up to non-global contributions, the jet function $J$ satisfies the RGE,
\bea
\mu \frac{\mathrm{d} J}{ \mathrm{d} \mu} = 
 \left( 2 \Gamma[\alpha_s] L   + 
\gamma_{J_q} 
\right)J \,, 
\eea
where $\Gamma$ is the cusp anomalous dimension, cf. e.g. \cite{Moch:2004pa}, 
\bea
\Gamma =  \sum_{n=0} \left( \frac{\alpha_s}{4\pi} \right)^{n+1} \Gamma_n  \,, 
\quad\quad \mbox{with} \quad\quad
\Gamma_0 = 4 C_F \,, \quad \quad  
\Gamma_1 = 4 C_F \left[ \left(\frac{67}{9}-\frac{\pi ^2}{3}\right) C_A - \frac{20}{9} T_F N_F \right] \,,
\eea
and $\gamma_{J_q}$ is the anomalous dimension of the quark-jet, 
\bea
\gamma_{J_q} = \sum_{n=0} \left( \frac{\alpha_s}{4\pi} \right)^{n+1} \gamma_n \,,  
\quad\quad \mbox{with} \quad\quad
\gamma_0 = 6 C_F \,. 
\eea
Using the fact that $ \mathrm{d} \log \mu =\beta^{-1}[\alpha_s] \,  \mathrm{d} \alpha_s  $, with $\beta[\alpha_s] = - 2 \alpha_s \sum_{n=0} \beta_n  \left(\frac{\alpha_s}{4\pi} \right)^{n+1}$, 
the RGE can be solved to find
\bea\label{eq:RGE}
J(\mu) = J(\mu = p_T R) \exp
\left[
2 \int_{\alpha_s(p_T R)}^{\alpha_s(\mu)} \mathrm{d} \alpha_s  
\frac{\Gamma[\alpha_s]}{\beta[\alpha_s]} \int_{\alpha_s(p_T R)}^{\alpha_s}
\frac{\mathrm{d}\alpha_s'}{\beta[\alpha_s']}
 \, +
 \int_{\alpha_s(p_T R)}^{\alpha_s(\mu)} \mathrm{d} \alpha_s \frac{\gamma_{J_q}[\alpha_s]}{\beta[\alpha_s]}
\right] \,.
\eea
Expanding the solution to ${\cal O}(\alpha_s^2)$, we find the logarithmic terms in the jet function at $\alpha_s^2$ contain
\bea\label{eq:log-rge}
J^{(2)} &=& \frac{\alpha_s^2}{(2\pi)^2} \left[
\frac{\Gamma_0^2}{8} L^4
+ \Gamma_0 \left(\frac{ \beta_0}{6} + \frac{\gamma_0}{4} \right)L^3
+ \frac{1}{8}\left(
2\beta_0 \gamma_0 + \gamma_0^2 + 4 \Gamma_0  J^{(1)}_{0,ren}
+ 2\Gamma_1
\right) L^2 
\right. \nn\\
&& \left.
+ \frac{1}{4}\left(
2 \gamma_0  J^{(1)}_{0,ren} 
+ 4 \beta_0  J^{(1)}_{0,ren}
+ \gamma_1 
\right) L
+ J^{(2)}_{0,ren}
\right] \,,
\eea
where $J^{(l)}_{0,ren}$ are the renormalized jet functions at NLO ($l=1$) and 
at NNLO ($l=2$) with $\mu = p_T R$ and $\epsilon \to 0$.

On the other hand, if we assume the bare jet function have the form
\bea
J_{bare} = 1 + Z_\alpha \frac{\alpha_s}{2\pi} e^{2\epsilon L} 
\left( \frac{\Gamma_0}{4\epsilon^2} + \frac{\gamma_0}{4\epsilon} 
+ J_0^{(1)} + J_1^{(1)}\epsilon 
\right)
+ \frac{\alpha_s^2}{(2\pi)^2} e^{4\epsilon L} \left(
\frac{J^{(2)}_{-4} }{\epsilon^4}
+\frac{J^{(2)}_{-3}}{\epsilon^3}
+\frac{J^{(2)}_{-2}}{\epsilon^2}
+\frac{J^{(2)}_{-1}}{\epsilon^1}
+J^{(2)}_0
\right) \,,
\eea
and hence the NLO renormalization factor of the jet function is 
\bea
Z_1 =  - \frac{\alpha_s}{2\pi} \left(
\frac{\Gamma_0}{4\epsilon^2} + \frac{\gamma_0}{4\epsilon} + 
\frac{\Gamma_0 \, L }{2\epsilon}
\right) \,.
\eea
It is straightforward to check that $J_{0}^{(1)} = J_{0,ren}^{(1)} $. 
From these, the logarithmic contributions can be obtained via
\bea\label{eq:log-pole}
J_{bare} + Z_1 + Z_1 J_{bare}^{(1)} &\supset& \frac{\alpha_s^2}{(2\pi)^2}
\left[  \left( \frac{32 J^{(2)}_{-4}}{3}  - \frac{5\Gamma_0^2}{24} \right)L^4
+ \left( \frac{32 J^{(2)}_{-3}}{3} - \frac{\beta_0\Gamma_0}{6} - \frac{5\gamma_0\Gamma_0}{12} \right)L^3
\right.
\nn\\
&&\left.
+ \left( 8 J^{(2)}_{-2} - \frac{1}{8}(\gamma_0 + 2 \beta_0) \gamma_0 - \frac{3}{2}\Gamma_0 J^{(1)}_0 \right)L^2
+ \left( 4 J^{(2)}_{-1} - \frac{1}{2}(\gamma_0 + 2 \beta_0) J^{(1)}_0 - \Gamma_0 J^{(1)}_1\right)L
\right] \,. \qquad
\eea
The coefficients of the $\epsilon$-poles $J_{-n}^{(2)}$ can be obtained 
by relating Eqs.~(\ref{eq:log-rge}) and (\ref{eq:log-pole}). 
The terms $J_{-n}^{(2)}$ with $n=4,3,2$ serve as a check, 
although due to the existence of the non-global logarithms, 
the $\epsilon^{-2}$-pole of $C_A C_F$ contribution will be affected and will deviate from the RGE prediction.
The term $J_{-1}^{(2)}$ allows for the extraction 
of the two-loop expression for the anomalous dimension of the quark-jet 
$\gamma_1$, which can be determined as 
\bea
\label{eq:g1}
\gamma_1  = {11.17(5)}  \, C_F^2 {-181.30(6)} \,  C_F C_A { -7.916(5)} \,  C_F N_F T_F \,. 
\eea

With quark-jet function computed to NNLO in Eq.~(\ref{eq:Jren}) 
and all anomalous dimensions known to the relevant order
we can also predict the logarithmic structure up to the
next-to-next-to-leading logarithm at ${\cal O}(\alpha_s^3)$ based on the naive RGE, which reads
\bea
J^{(3)} &=& 
\frac{\alpha_s^3}{(2\pi)^3} \Bigg\{
\frac{\Gamma_0^3}{48} L^6
+ \frac{1}{48} \left( 4\beta_0 + 3\gamma_0 \right) \Gamma_0^2 \, L^5
+ \frac{1}{48}\Gamma_0 \left(
4\beta_0^2 + 10 \beta_0\gamma_0 + 3\gamma_0^2 + 6 J^{(1)}_{0,ren} \Gamma_0 + 6\Gamma_1
\right)L^4 \nn \\
& &+  \frac{1}{48} \left(
8\beta_0^2\gamma_0 + 6\beta_0\gamma_0^2 + \gamma_0^3 +4\beta_1\Gamma_0 + 6\Gamma_0\gamma_1
+ 6\gamma_0(2J^{(1)}_{0,ren} \Gamma_0 + \Gamma_1)
+ 8\beta_0(4J^{(1)}_{0,ren} \Gamma_0 + \Gamma_1) 
\right) L^3\nn \\
& &+ \left(
J^{(1)}_{0,ren} \beta_0^2 + \frac{3}{4} J^{(1)}_{0,ren} \beta_0\gamma_0
+ \frac{\gamma_0\beta_1}{8} 
+ \frac{\gamma_0^2}{8} J^{(1)}_{0,ren} \,
+ \frac{\Gamma_0}{2} J^{(2)}_{0,ren} \,
+ \frac{\gamma_1 \beta_0}{4}
+ \frac{\gamma_0\gamma_1}{8}
+ \frac{\Gamma_1}{4} \, J^{(1)}_{0,ren}
+ \frac{\Gamma_2}{8}
\right)L^2 
\nn \\
&& + {\cal O}(L) \Bigg\}
\, ,
\eea
where sub-leading logarithmic terms require additional inputs from higher order calculations.

Finally, we note that in dimensional regularization the $\epsilon$-poles of the quark-jet function 
display the a pattern of exponentiation analogous to the quark form-factor,
cf. e.g.~\cite{Moch:2005id}, 
with the cusp anomalous dimensions determining the leading poles 
and the anomalous dimension of the quark-jet $\gamma_{J_q}$ 
governing the sub-leading ones.

\section{Analytic results}
\label{app:C}
The analytic computation of some of the phase space integrals associated with Eq.~(\ref{eq:inclusive})
\bea
d\Phi_{incl.} \equiv \sum_{i={\text I}}^{\text{III}} d \Phi_{3,sub.}^{i}   \times (p_T R)^{-4}
=
d \Phi_3  \times (p_T R)^{-4} \times  \theta( \tilde{s}_{jk} <1 )
\theta\left(  \tilde{s}_{ik} < 1 \right) \, \theta(z_i < z_j < z_k )  \,,  
\eea
is straightforward, specifically for integrals like
\bea
(a)=  d\Phi_{incl.} \, \frac{1}{\tilde{s}_{ik} \tilde{s}_{jk}} \,, \quad  \quad
\text{and} \quad \quad
(b)=d\Phi_{incl.}  \, \frac{1}{\tilde{s}_{ik} \tilde{s}_{ij}} \,.
\eea
For $(a)$ we directly obtain 
\bea
\label{eq:analytic-a}
(a) &=& 
    2^{2-4\epsilon} (p_T R)^{-4\epsilon} 
\frac{ \int_0^1  d z_i dz_j  }{(4\pi)^{5-2\epsilon} \Gamma(1-2\epsilon) }
   (z_iz_jz_k)^{1 - 2\epsilon}      \,
  \int  d\tilde{s}_{ik} d\tilde{s}_{jk}  
      \left(   \tilde{s}_{ik} \tilde{s}_{jk}  \right)^{-1-\epsilon }  
 \int_{0}^{1} dt  \, t^{-\frac{1}{2} - \epsilon}   
\left(
   1-t
\right)^{-\frac{1}{2} - \epsilon}   \nn \\
   &=&  
\frac{(p_T R)^{-4\epsilon}   }{\epsilon^2 \Gamma (1-\epsilon)^2}
\frac{     \int_{0}^{1}    d z_i dz_j  }{(4\pi)^{4-2\epsilon}   }
   (z_iz_jz_k)^{1 - 2\epsilon}  \theta(z_i<z_j<z_k)
\,,
\eea
while for $(b)$, we find
\bea
(b) &=& 
    2^{2-4\epsilon} (p_T R)^{-4\epsilon} 
\frac{    \int_{0}^{1}    d z_i dz_j   }{(4\pi)^{5-2\epsilon} \Gamma(1-2\epsilon) }
   (z_iz_jz_k)^{1 - 2\epsilon}       \nn \\
&&\times     
\int  d\tilde{s}_{ik} d\tilde{s}_{jk}
    \tilde{s}_{ik}^{-1-\epsilon }  \tilde{s}_{jk}^{-\epsilon}
  \int_{0}^{1} dt
\left(
(\sqrt{ \tilde{s}_{jk} } -\sqrt{ \tilde{s}_{ik}})^2+4\sqrt{ \tilde{s}_{jk}  \tilde{s}_{ik}} t 
\right)^{-1}
\, t^{-\frac{1}{2} - \epsilon}   
\left(
   1-t
\right)^{-\frac{1}{2} - \epsilon}   \nn \\
&=&
2^{2-4\epsilon} (p_T R)^{-4\epsilon} \,     
\frac{\Gamma\left(\frac{1}{2}-\epsilon\right)^2}
{\Gamma(1-2\epsilon)}\, 
\frac{      \int_{0}^{1}     d z_i dz_j  }{(4\pi)^{5-2\epsilon} \Gamma(1-2\epsilon) }
   (z_iz_jz_k)^{1 - 2\epsilon}       \nn \\
&&\times
\int  d\tilde{s}_{ik} d\tilde{s}_{jk}
  \tilde{s}_{ik}^{-1-\epsilon }  \tilde{s}_{jk}^{-\epsilon}
\,
\left(  \tilde{s}_{ik} + \tilde{s}_{jk} -2 \sqrt{ \tilde{s}_{ik}  \tilde{s}_{jk} }\right)^{-1}
{}_2F_1\left[ 1,\frac{1}{2}-\epsilon,1-2\epsilon,- \frac{4 \sqrt{ \tilde{s}_{ik}  \tilde{s}_{jk}}} {   
\tilde{s}_{ik} + \tilde{s}_{jk} -2 \sqrt{ \tilde{s}_{ik}  \tilde{s}_{jk} }
} \right] \,. 
\eea
By splitting up the integral into two regions, 
$\tilde{s}_{ik} \le \tilde{s}_{jk}$ and $\tilde{s}_{ik} > \tilde{s}_{jk}$, 
we have for 
$\tilde{s}_{ik} \le \tilde{s}_{jk}$, we let $\tilde{s}_{ik} = \tilde{s}_{jk} u^2 $,
\bea
(b_1) &=&
2^{2-4\epsilon} (p_T R)^{-4\epsilon} 
\frac{\Gamma\left(\frac{1}{2}-\epsilon\right)^2}
{\Gamma(1-2\epsilon)}\
\frac{     \int_{0}^{1}     d z_i dz_j  }{(4\pi)^{5-2\epsilon} \Gamma(1-2\epsilon) }
   (z_iz_jz_k)^{1 - 2\epsilon}       \nn \\
&&\times      
2 \int_0^1 d\tilde{s}_{jk}  \tilde{s}_{jk}^{-1-2\epsilon} \,
\int_0^1 du
u^{-1-2\epsilon } 
\,
\left(    1- u  \right)^{-2}
{}_2F_1\left[ 1,\frac{1}{2}-\epsilon,1-2\epsilon,\frac{- 4 u }{   
(1-u)^2
} \right]
   \nn \\
   &=&
2^{2-4\epsilon} (p_T R)^{-4\epsilon} \,
\frac{\Gamma\left(\frac{1}{2}-\epsilon\right)^2}
{\Gamma(1-2\epsilon)}\,
\frac{     \int_{0}^{1}     d z_i dz_j   }{(4\pi)^{5-2\epsilon} \Gamma(1-2\epsilon) }
   (z_iz_jz_k)^{1 - 2\epsilon}       \nn \\
&&\times      
2  \int_0^1 d\tilde{s}_{jk}  \, 
 \tilde{s}_{jk}^{-1-2\epsilon}
\int_0^1 du \, 
  u^{-1-2\epsilon } 
\,
{}_2F_1\left[ 1,1+\epsilon,1-\epsilon,u^2
 \right]
   \nn \\
      &=&
(p_T R)^{-4\epsilon} 
\frac{      \int_{0}^{1}     d z_i dz_j   }{(4\pi)^{4-2\epsilon} \Gamma(1-2\epsilon) }
   (z_iz_jz_k)^{1 - 2\epsilon}  
\left( 
\frac{3}{4 \epsilon^2} -\frac{\pi ^2}{24}
-\frac{5  \zeta_3}{2} \epsilon
-\frac{3 \pi ^4}{32}  \epsilon^2
\right) \,, 
\eea
where we have used
\bea
{}_2F_1\left[ 1,\frac{1}{2}-\epsilon,1-2\epsilon,\frac{- 4 u }{   
(1-u)^2
} \right]
=(1-u)^2
{}_2F_1\left[ 1,1+\epsilon,1-\epsilon,u^2 \right] \,.
\eea
For $\tilde{s}_{jk} < \tilde{s}_{ik}$, we let $\tilde{s}_{jk} = u^2 \tilde{s}_{ik}$ and obtain 
\bea
(b_2) &=&
 (p_T R)^{-4\epsilon} 
\frac{       \int_{0}^{1}     d z_i dz_j   }{(4\pi)^{4-2\epsilon} \Gamma(1-2\epsilon) }
   (z_iz_jz_k)^{1 - 2\epsilon} 
\left(
\frac{1}{4\epsilon^2} + \frac{\pi^2}{24} 
+\frac{\zeta_3}{2} \epsilon + \frac{19\pi^4}{1440} \epsilon^2
\right)\,.
\eea
In summary, this gives us
\bea
\label{eq:analytic-b}
(b) = (b_1)+ (b_2) =  \frac{(p_T R)^{-4\epsilon} }{\Gamma(1-2\epsilon)}
\frac{       \int_{0}^{1}     d z_i dz_j   }{(4\pi)^{4-2\epsilon}  }
    (z_iz_jz_k)^{1 - 2\epsilon} \, \theta(z_i<z_j<z_k) 
  \left(
  \frac{1}{ \epsilon^2} 
  - 2 \zeta_3 \epsilon
  - \frac{29 \pi^4}{360}  \epsilon^2
  \right) \,. 
\eea

The results of $(a)$ and $(b)$ in Eqs.~(\ref{eq:analytic-a}) and (\ref{eq:analytic-b})
can be used to validate the numerical calculations. 
For instance, for the term proportional to $\frac{1}{s_{12}s_{13}}$ in the
splitting function $P_{{\bar q}_1q_2q_3}^{({\rm id})}$,
\bea
P^{({\rm id})}_{{\bar q}_1q_2q_3}\left[\frac{1}{s_{12}s_{13}} \right] = 
4(4\pi)^2\alpha_s^2 \left( \frac{\mu^2 e^{\gamma_E}}{4\pi} \right)^{2\epsilon}
\frac{-1}{2s_{12} s_{13}} z_1 \left[
\frac{1+z_1^2}{(1-z_2)(1-z_3)} -\epsilon\left( 1+2\frac{1-z_2}{1-z_3}\right) - \epsilon^2 
\right]  \,, 
\eea
we have the following result for $z_1<z_2<z_3$, 
by using $(b)$ in Eq.~(\ref{eq:analytic-b}) and after integrating over $z_1$ and $z_2$, 
\bea
\frac{\alpha_s^2}{(2\pi)^2} \, e^{4\epsilon L}\,
\left(
\frac{-0.189353}{\epsilon^2}
-\frac{1.70334}{\epsilon}
-8.02839
\right) \,,
\eea
which agrees with the numerical calculation, 
\bea
 \frac{\alpha_s^2}{4\pi^2} \, e^{4\epsilon L}
\left( \frac{-0.18935(1)   }{\epsilon^2}
- \frac{1.70335(7)  }{\epsilon}
-8.0283(5)   \right) \,.
\eea
For $z_2<z_3<z_1$, using $(a)$ in Eq.~(\ref{eq:analytic-a}) and integrating over $z_2$ and $z_3$, we find
\bea
 \frac{\alpha_s^2}{4\pi^2} \, e^{4\epsilon L}
\left(
\frac{-0.0911635}{\epsilon^2}
-\frac{0.668211}{\epsilon}
-2.3708
\right) \,,
\eea
in agreement with the numerical calculation, which returns
\bea
 \frac{\alpha_s^2}{4\pi^2} e^{4\epsilon L}
\left( \frac{-0.091162(2)  }{\epsilon^2}
- \frac{0.66821(2) }{\epsilon}
-2.3709(1)   \right) \,.
\eea
The same check can be done for the splitting function $P_{g_1g_2q_3}$.
For instance for the contribution proportional to $\frac{1}{s_{12}s_{13}}+(1\leftrightarrow 2)$ 
in the non-Abelian term $P_{g_1g_2q_3}^{({\rm nab})}$
\bea
P_{g_1g_2q_3}^{({\rm nab})}\left[
\frac{1}{s_{12}s_{13}}+(1\leftrightarrow 2)
\right]
&= &
4(4\pi)^2\alpha_s^2 \left(\frac{\mu^2 e^{\gamma_E}}{4\pi} \right)^{2\epsilon} \,
\Bigg\{
\, \frac{1}{2s_{12}s_{13}} 
\left[
\frac{2z_3+(1-\epsilon)(1-z_3)^2}{z_2}
+ \frac{2(1-z_2)+(1-\epsilon )z_2^2}{1-z_3}
\right]  \Bigg\} 
  \nn\\
& &
+(1\leftrightarrow 2) \,,
\eea
we find for $z_1< z_2 <z_3$, 
\bea
\frac{ \alpha_s^2 }{4\pi^2} e^{4\epsilon L}
\left(
\frac{3}{8\epsilon^4}
+ \frac{21+48 \log 2}{32\epsilon^3}
+ \frac{2.23062}{\epsilon^2}
- \frac{1.7911}{\epsilon}
+0.703757
\right) \,, 
\eea
again in complete agreement with the numerical calculations
\bea
\frac{ \alpha_s^2 }{4\pi^2} e^{4\epsilon L}
\left(
\frac{0.375 }{\epsilon^4}
+ \frac{1.69596(5)   }{\epsilon^3}
+ \frac{2.2305(2)}{\epsilon^2}
- \frac{1.791(8) }{\epsilon}
+0.703(3) 
\right) \,.  
\eea
Similar other checks have also been performed but are not listed here.

\end{widetext}

\bibliographystyle{h-physrev3.bst}
\bibliography{nnlo-jet_v2}

\end{document}